\documentclass[12pt,twoside]{article}   
\usepackage[super,sort,comma]{natbib}

\usepackage{fancyhdr}		

\newcommand{\captionv}[3]{\begin{center}\parbox{#1cm}{\caption[#2]{{\sf #3}}}
        \end{center}}

\usepackage[section]{placeins}   %

\usepackage{graphicx}

\makeatletter \renewcommand\@biblabel[1]{$^{#1}$} \makeatother
\newcommand{\cen}[1]{\begin{center} #1 \end{center}}

\usepackage{hyperref}
\hypersetup{ colorlinks,
    citecolor=blue,
    filecolor=blue,
    linkcolor=blue,
    urlcolor=blue
}

\usepackage{xcolor}

\definecolor{gray}{rgb}{0.6,0.6,0.6}
\definecolor{red}{rgb}{0.85,0,0}
\definecolor{green}{rgb}{0,0.85,0}
\definecolor{blue}{rgb}{0,0,0.85}
\definecolor{beige}{rgb}{0.92,0.87,0.78}
\usepackage[all]{hypcap}    

\usepackage{makecell}
\usepackage{algorithm}
\usepackage{algpseudocode}
\usepackage{multirow}
\usepackage{amsmath}
\usepackage{url}
\usepackage[normalem]{ulem}
\usepackage{xcolor}
\usepackage{soul}

\begin{document}

\cen{\sf {\Large {\bfseries Federated prediction for scalable and privacy-preserved knowledge-based planning in radiotherapy} \\  
\vspace*{10mm}
Jingyun Chen$^1$, David Horowitz$^{1,2}$ and Yading Yuan$^{1,3}$} \\
$^1$Department of Radiation Oncology, Columbia University Irving Medical Center, New York, NY, USA \\
$^2$Herbert Irving Comprehensive Cancer Center, Columbia University, New York, NY, USA\\
$^3$Data Science Institute, Columbia University, New York, NY, USA
}

\pagenumbering{roman}
\setcounter{page}{1}
\pagestyle{plain}
\noindent\textbf{Corresponding Author: }\\
Yading Yuan, PhD\\
Department of Radiation Oncology\\
Columbia University Irving Medical Center\\
622 W 168th Street, CHONY-B11\\
New York, NY, 10032\\
Email: yading.yuan@columbia.edu
\newpage
\begin{abstract}
\noindent {\bf Background:} Deep learning has potential to improve the efficiency and consistency of radiation therapy planning, but clinical adoption is hindered by the limited model generalizability due to data scarcity and heterogeneity among institutions. Although aggregating data from different institutions could alleviate this problem, data sharing is a practical challenge due to concerns about patient data privacy and other technical obstacles.\\ 
{\bf Purpose:} This work aims to address this dilemma by developing FedKBP+, a comprehensive federated learning (FL) platform for predictive tasks in real-world applications in radiotherapy treatment planning.\\
{\bf Methods:} We implemented a unified communication stack based on Google Remote Procedure Call (gRPC) to support communication between participants whether located on the same workstation or distributed across multiple workstations. In addition to supporting the centralized FL strategies commonly available in existing open-source frameworks, FedKBP+ also provides a fully decentralized FL model where participants directly exchange model weights to each other through Peer-to-Peer communication. We evaluated FedKBP+ on three predictive tasks using scale-attention network (SA-Net) as the predictive model.\\
{\bf Results:} Using 340 cases (training: 200; validation: 40; testing: 100) from the OpenKBP Challenge, a 3D dose prediction model trained with FedAvg algorithm outperformed the model trained on the local data, and achieved predictive accuracy comparable to that of a centrally trained model using pooled data in both independent and identically distributed (IID) and non-IID settings. We further evaluated the performance of FedKBP+ against NVFlare on the task of brain tumor segmentation using 227 cases from eight sites in the 2021 BraTS challenge dataset (training: 152; validation: 27; testing: 48). FedKBP+ surpassed the NVFlare framework in both accuracy and training efficiency. FedAvg with FedKBP+ achieved a DSC of 92.38\% in 5.92 hours versus NVFlare’s 90.75\% in 7.78 hours; FedProx with FedKBP+ reached 91.98\% in 6.03 hours against NVFlare's 90.37\% in 7.80 hours. Finally, in a robustness study of FedKBP+ for organ segmentation using 384 cases from five sites in the PanSeg dataset (training: 269; validation: 39; testing: 76), Gossip Contrastive Mutual Learning - a novel decentralized FL algorithm - demonstrated strong resilience to site failure without significant accuracy loss even when up to 40\% of sites randomly dropped out at each round of federated training ($p=0.9097$).\\
{\bf Conclusions:} Our results demonstrate that FedKBP+ is highly effective, efficient and robust, showing great potential as a federated learning platform for radiation therapy.
\end{abstract}


\newpage
\setlength{\baselineskip}{0.7cm}      

\pagenumbering{arabic}
\setcounter{page}{1}
\pagestyle{fancy}
\section{Introduction}

In recent years, deep learning (DL)-based predictive modeling has emerged as a pivotal technology in radiation therapy (RT), significantly augmenting the efficiency and standardization of various steps in treatment planning workflows, including (1) delineation of Organs at Risk (OAR) \cite{Pan2023, Li2023, Tian2023, Lin2021, Liu2020}, (2) segmentation of tumor targets \cite{Baek2025, Jiang2025, Dong2024, Li2022c, Zhuge2017}, and (3) Knowledge-Based Planning (KBP) \cite{Wu2009}, particularly the three-dimensional (3D) dose prediction \cite{Yuan2018, Babier2021, Chen2024a, He2025, Fransson2024}. However, these models often struggle to generalize outside the developing institutions due to data heterogeneity stemming from inter-institutional variations in patient demographics, imaging protocols, annotation standards, treatment methodologies, and other contextual factors. Meanwhile,  there are large varieties in training these models from model architecture, pre- and post-processing steps, data augmentation and optimization procedures. These bring additional challenges for small hospitals with limited resources to optimize these models to meet their unique needs. While centralizing data from multiple institutions, e.g. OpenKBP study \cite{Babier2021}, can potentially improve model generalizability, such centralized training strategies are rarely practicable in clinical settings due to stringent regulatory governance on patient  privacy that preclude institutions from sharing medical data, even when such collaboration could yield more institutionally adaptive models.

Alternatively, Federated Learning (FL) \cite{McMahan2017, Rieke2020} presents a compelling paradigm for collaborative model development that removes the need to exchange sensitive patient data. In the FL framework, each participating institution independently trains a local model on its proprietary dataset and subsequently shares only model updates, rather than raw data, to collaborators. This approach allows institutions to benefit from external data—often difficult and time-consuming to curate independently—while preserving data privacy and minimizing the risk of exposing sensitive information.

Based on the communication topology employed for model update exchange, FL algorithms can be broadly classified into centralized and decentralized FL  \cite{Rieke2020}. In centralized FL, the participating institutions transmit model updates to a central server for aggregating them into a global model, while in decentralized FL, updates are shared directly among institutions without the involvement of an aggregation server. Owing to its inherent privacy-preserving design, FL has gained considerable attention in medical domains, where stringent data protection regulations frequently limit cross-institutional data sharing \cite{Shen2023, Feng2023, Dong2023}. A prominent example is the large-scale international initiative involving 71 institutions across six continents with over 6,000 patients, in which FL-based training achieved a 20\% performance improvement over a conventional baseline for automated brain tumor segmentation using multi-parametric MRI scans \cite{Pati2022}.

Nevertheless, the deployment of FL in real-world scenarios is encumbered by several technical challenges, including communication security and scalability of the distributed system \cite{Rieke2020}. To mitigate these issues and promote broader FL adoption, a number of open-source frameworks have been introduced, such as OpenFL \cite{Foley2022}, NVIDIA FLARE (NVFlare) \cite{Roth2021}, and Flower \cite{Beutel2020}. 

OpenFL, developed by Intel, is a Python-based FL library \cite{Foley2022} that was used in one of the largest global healthcare FL initiatives \cite{Pati2022}. Its design primarly focuses on centralized paradigms such as FedAvg \cite{McMahan2017} and FedProx \cite{Li2018}. A separate and unrelated project also named OpenFL, which uses blockchain to support decentralized FL \cite{Wahrsttter2024}, has yet to demonstrate real-world deployments and is not currently considered suitable for medical federations.

NVFlare is an open source and extensible FL platform developed by NVIDIA, aimed at supporting domain-agnostic and customizable workflows \cite{Roth2022}. It includes a Client-Controlled Workflow mechanism \cite{URL_NVFlare}, which enables implementations of decentralized FL strategies such as Cyclic Learning \cite{Yu2022}, Swarm Learning \cite{Gao2023}, etc. Nevertheless, NVFlare fundamentally operates on a client-server architecture, which constrains its support for fully decentralized FL.

Flower, developed by the University of Cambridge, offers robust support for large-scale experimentation and heterogeneous FL environments \cite{Beutel2020}. Although it offers partial support for decentralized FL, including a use case involving Cyclic Learning \cite{URL_Flower}, its model exchange mechanism still relies on a central server to sequentially relay models among clients.

Although these platforms have enhanced the accessibility and adaptability of FL implementations, most are predominantly optimized for centralized FL and offer limited built-in support for decentralized FL. Reliance on central servers in existing FL frameworks introduces inherent architectural vulnerabilities, including single point of failure and increased susceptibility to cyberattacks during deployment. In addition, most current FL frameworks offer limited support or task-specific customization for radiation therapy applications, which hinders their ability to address the unique clinical, anatomical, and dosimetric complexities associated with RT workflows.

To overcome these limitations, we introduce FedKBP+, a comprehensive FL framework that includes both centralized and decentralized FL strategies.  It is specifically engineered to address the unique challenges of real-world FL scenarios in RT treatment planning, supporting key predictive tasks including OAR delineation, tumor segmentation, and dose prediction. Table \ref{tab_1} compares these FL frameworks.

\begin{table}[htbp]
\begin{center}
\captionv{16}{}{Compare of FedKBP+ and other open-source FL frameworks
\label{tab_1}
\vspace*{2ex}
}
\begin{tabular} {|l|c|c|c|c|}
\hline
        &               &               &               &        \\
       Frameworks &  \thead{Support \\ Centralized FL}  & \thead{Support \\ Decentralized FL} & \thead{Support direct P2P \\ model exchange}  & \thead{RT-related \\ implementations} \\
\hline
        &               &               &               &        \\
OpenFL \cite{Foley2022}     & Fully	& Not yet	& Not yet	&  No         \\
&&&&  \vspace{-2mm}\\
NVFlare \cite{Roth2021}     & Fully	& Partially	& Not yet	&   No        \\
&&&&  \vspace{-2mm}\\
Flower \cite{Beutel2020}     & Fully & Partially	& Not yet	&  No         \\
&&&&  \vspace{-2mm}\\
FedKBP+ (ours)      & Fully	& Fully	& Fully	& Yes          \\
        &               &               &               &        \\
\hline
\end{tabular}
\end{center}
\end{table}

The primary contributions of this work are as follows.

\begin{figure}[ht]
   \begin{center}
   \includegraphics[width=15cm]{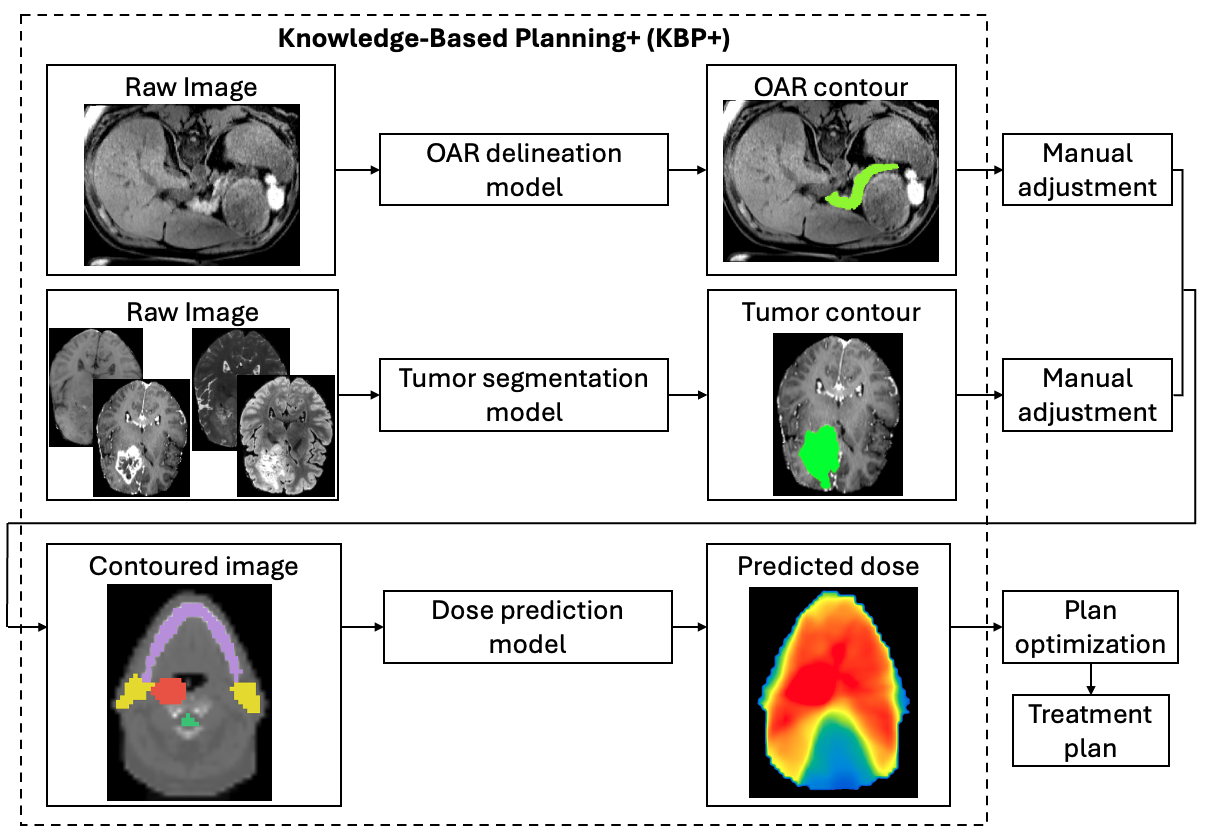}
   \captionv{16}{}
   {Workflow of the three predictive tasks in the Knowledge-Based Planning+ (KBP+) framework
   \label{fig_kbp+} 
    }
    \end{center}
\end{figure}

(1) We extend the scope of KBP, which has traditionally focused on the prediction of dose-volume histograms (DVH) or 3D dose distributions, to a broader and more integrative framework termed \textbf{KBP+}. As illustrated in Figure \ref{fig_kbp+}, KBP+ unifies the three main predictive tasks in RT planning: tumor segmentation, OAR segmentation, and dose prediction. This allows us to integrate various tasks into a unified machine learning pipeline to facilitate collaborative mode training across different centers through federated learning.

\begin{figure}[ht]
   \begin{center}
   \includegraphics[width=13cm]{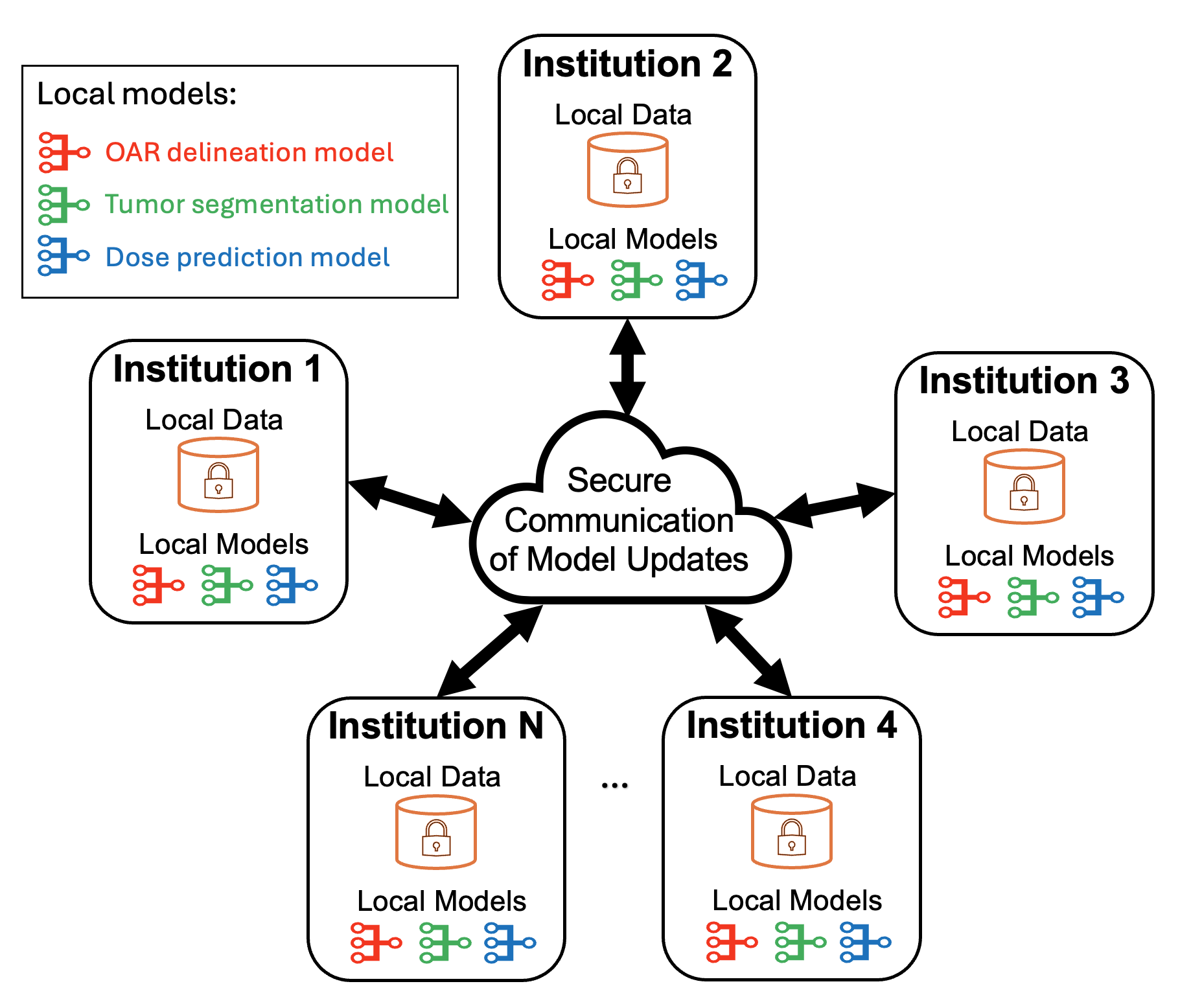}
   \captionv{16}{}
   {FedKBP+: the FL framework for the predictive models in KBP+
   \label{fig_fedkbp+} 
    }
    \end{center}
\end{figure}

(2) We present \textbf{FedKBP+} (Figure \ref{fig_fedkbp+}), a comprehensive FL framework designed to support the full range of predictive tasks in KBP+. We perform a comprehensive evaluation of FedKBP+'s performance across these tasks, benchmarking it against other FL framework and conventional non-FL training approaches.

(3) We will provide FedKBP+ as an open source package, including use cases for each predictive task in FedKBP+ and a backbone model (SA-Net) that achieves competitive performance in all of these tasks. Upon acceptance of this paper, the open source code will be released on \url{https://github.com/CUMC-Yuan-Lab/FedKBP_plus}.

\section{Methods}
In this section, we present the details of the FedKBP+ framework, covering the core FL components, centralized and decentralized FL algorithms, the predictive model, and the communication protocols.

\subsection{Preliminaries}
In this paper, the collaborating institutions in FL are referred to as sites. Each site maintains a local file system (e.g., a hard drive) containing its own data and model. During FL, each site executes a pre-installed program, referred to as the FL script, on its local machine. This script initiates the FL engine on a GPU or a CPU to perform various tasks such as model training, parameter sharing, and aggregation. The scheduling and coordination of these tasks across sites are governed by specific FL algorithms, such as FedAvg \cite{McMahan2017}, FedProx \cite{Li2018}, and GCML \cite{Chen2025}.

Throughout the remainder of this paper, the total number of participating sites is denoted as $N$. For site $i\in [1,2,...,N]$, we denote the local model weights as $w_i$, the number of local cases as $m_i$, and the learning rate as $\eta_i$. The total number of cases across all sites is represented by $m$, where $m=\sum_{i=1}^{N}m_i$. The weights of the aggregated (i.e., global) model are denoted by $w$.

\subsection{FL workflows in FedKBP+}
\begin{figure}[ht]
   \begin{center}
   \includegraphics[width=15cm]{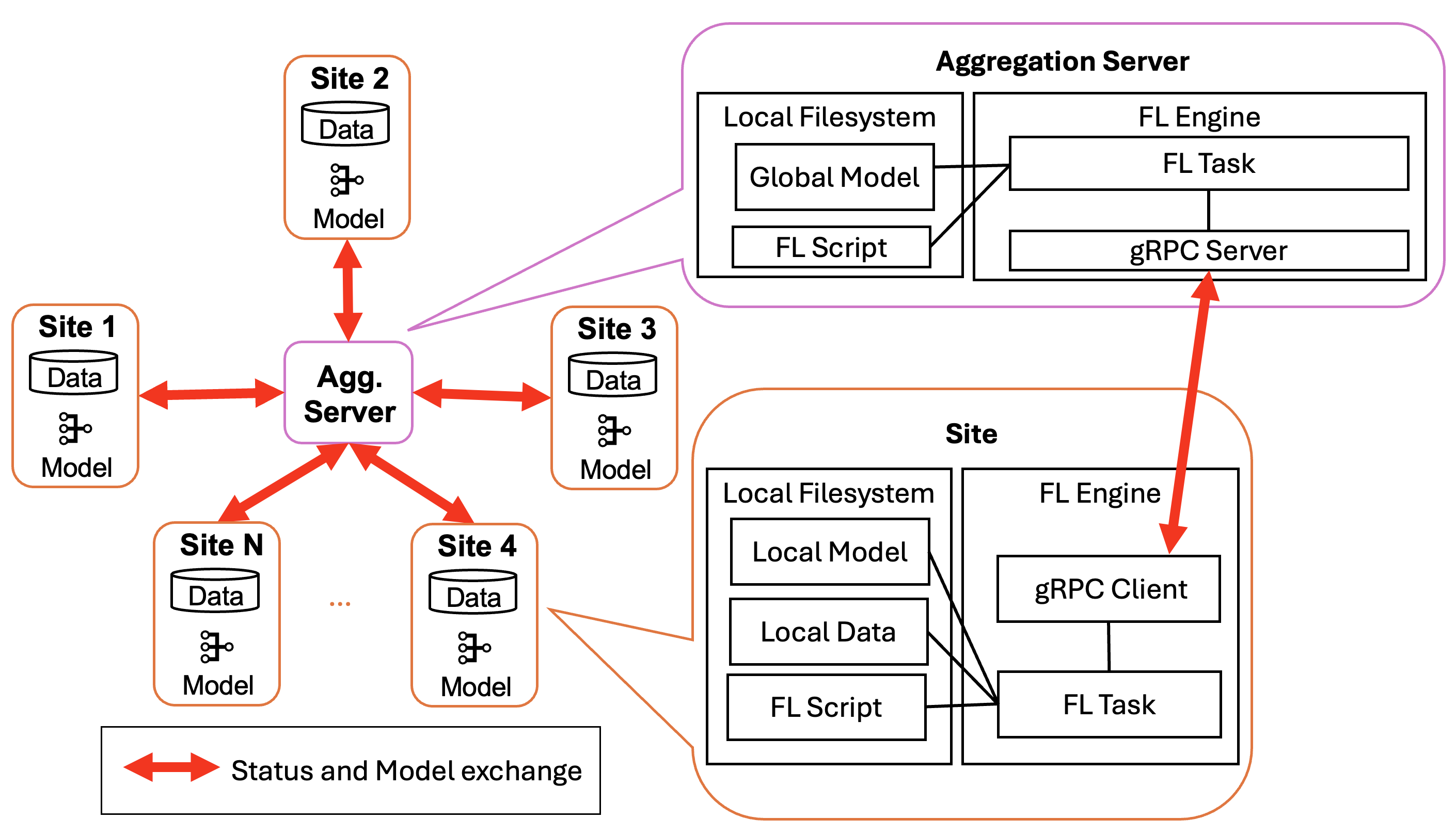}
   \captionv{16}{}
   {Centralized FL in FedKBP+ framework.
   \label{fig_centralized_fl} 
    }
    \end{center}
\end{figure}

In centralized FL, a central aggregation server is responsible for maintaining the global model. During each FL round, participating sites upload their local model updates to this server. The server aggregates these updates to refine the global model, which is then broadcast back to all sites. Each site then replaces its local model with the updated global model and proceeds to the next round of local training, as shown in Figure \ref{fig_centralized_fl}.

\begin{figure}[ht]
   \begin{center}
   \includegraphics[width=15cm]{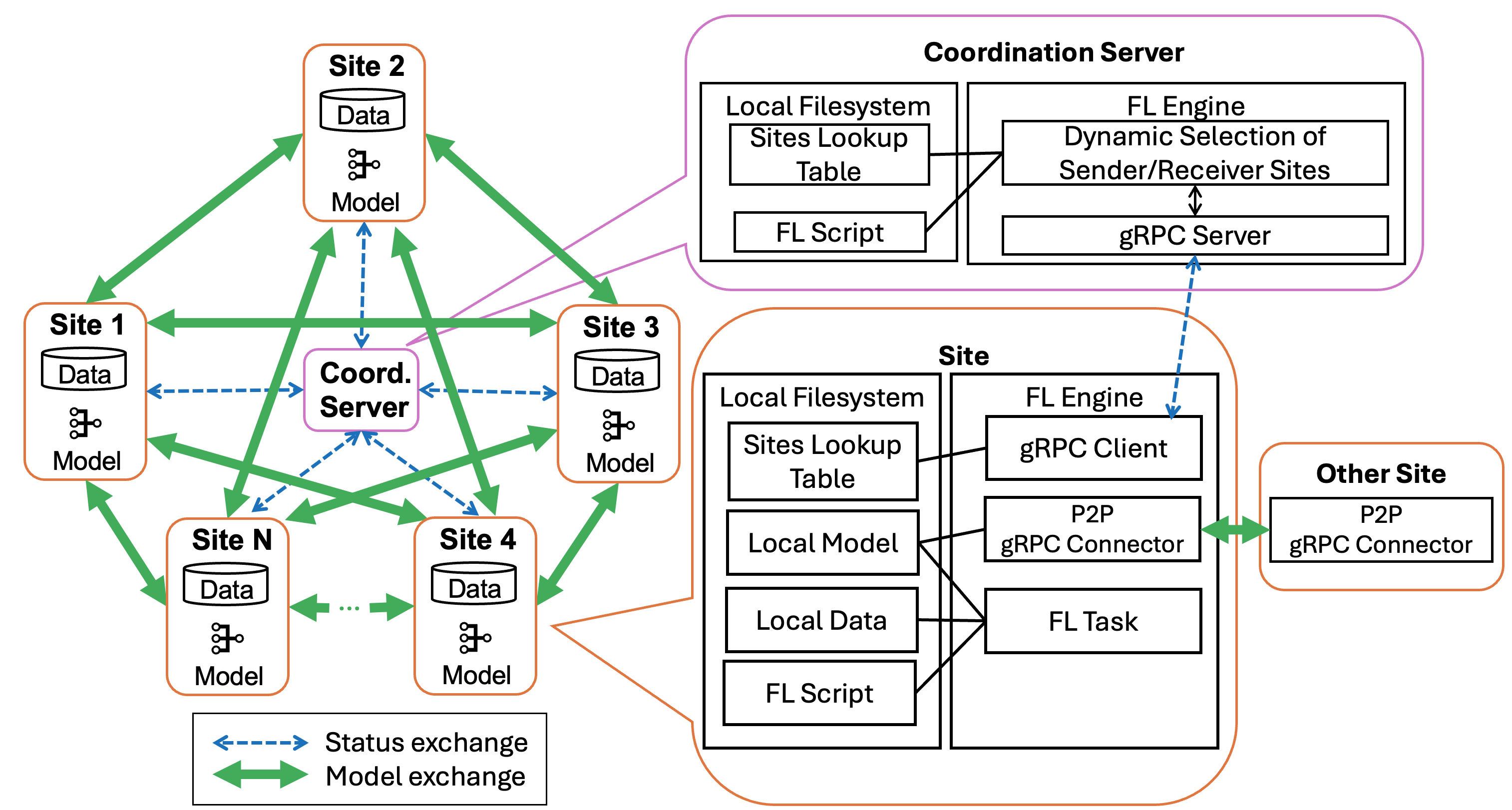}
   \captionv{16}{}
   {Decentralized FL in FedKBP+ framework.
   \label{fig_decentralized_fl} 
    }
    \end{center}
\end{figure}

In decentralized FL, a coordination server is established. However, unlike the aggregation server in centralized FL, it neither receives model updates from sites nor maintains a global model. Instead, its role is to track and manage site metadata such as IP address and port, FL participation status (active or dropped out), and each site's role in the model exchange process (sender or receiver). At the beginning of each FL round, the coordination server broadcasts this updated metadata to all participating sites. Based on their assigned roles, the sites then engage in P2P communication, either sending their local models to or receiving models from designated peer sites, as shown in Figure \ref{fig_decentralized_fl}.

\subsection{Predictive model}
While the predictive model in FedKBP+ is fully customizable, its architectural design plays a crucial role in determining the accuracy of predictions. In this paper, we employed the Scale Attention Network (SA-Net) \cite{Yuan2021,Yuan2021a} as the predictive model, utilizing task-specific loss functions tailored to each application.

\begin{figure}[ht]
   \begin{center}
   \includegraphics[width=16cm]{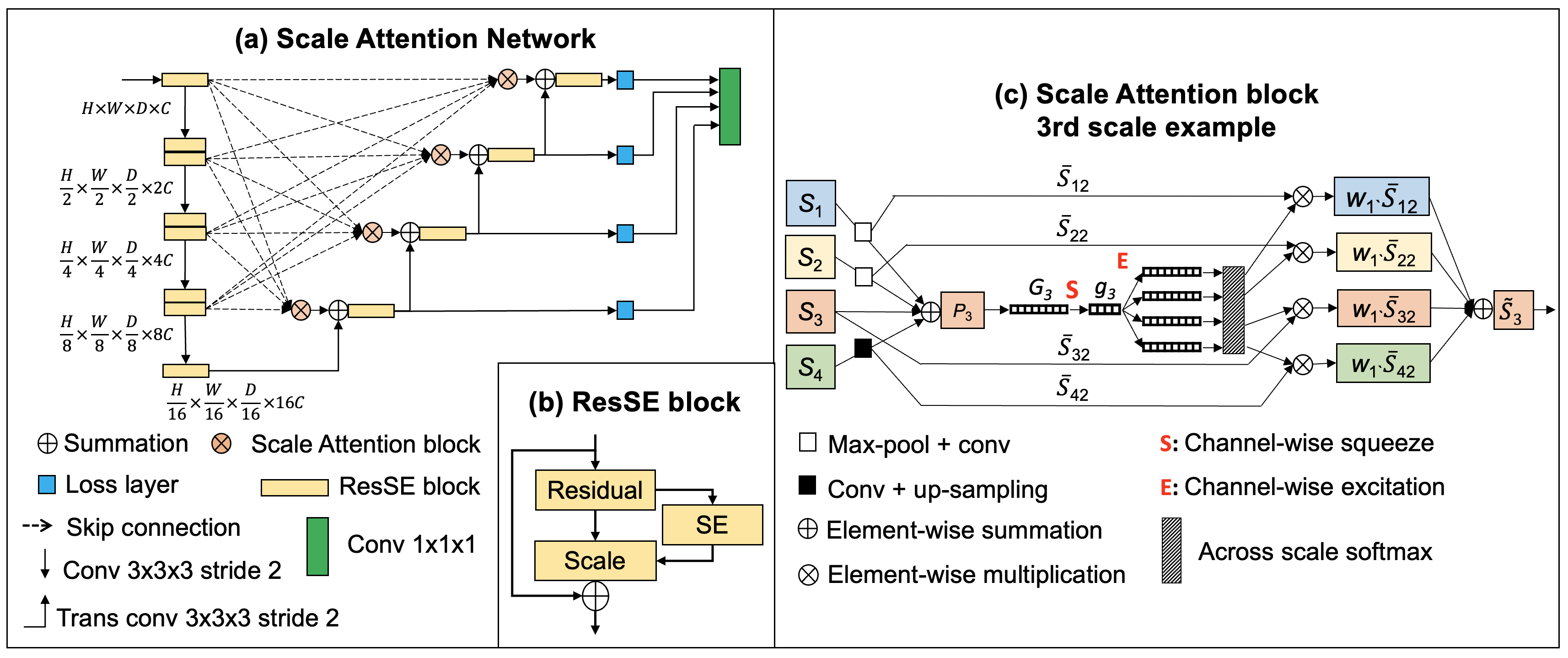}
   \captionv{16}{}
   {Architecture of SA-Net(a) including ResSE block(b) and Scale Attention block(c).
   \label{fig_sanet}
    }
    \end{center}
\end{figure}

As shown in Figure \ref{fig_sanet}(a), SA-Net adopts a classic encoder–decoder architecture. The encoder is built upon the ResNet framework \cite{He2016}, and its representational capacity is enhanced by integrating a squeeze-and-excitation (SE) module \cite{Hu2018} into each residual block, resulting in ResSE blocks (Figure \ref{fig_sanet}(b)). The decoder mirrors the encoder’s structure in reverse, but contains only a single ResSE block per level. Upsampled feature maps in the decoder are fused with the outputs from the scale attention block using element-wise summation—rather than concatenation—to reduce parameter overhead while preserving information. To facilitate effective training and improve gradient flow, deep supervision is applied across all intermediate scales. This acts as a regularization mechanism and enhances model convergence. An example scale attention block (for the third decoding level) is depicted in Figure \ref{fig_sanet}(c). First, encoder outputs from various scales are reshaped to a uniform spatial resolution. These feature maps are then summed and passed through a global average pooling layer, followed by the squeeze-and-excitation operation. The resulting outputs are normalized using a softmax function to generate scale-specific weight vectors for each channel. The final output is obtained as a weighted sum of these multi-scale vectors.

SA-Net’s dynamic scale attention mechanism effectively fuses fine-grained spatial details with high-level contextual semantics across scales, enabling robust multi-scale feature integration. This architecture has demonstrated strong performance across multiple applications, including 3D dose prediction \cite{Yuan2017}, tumor target segmentation \cite{Yuan2020,Yuan2021,Yuan2022}, and organ delineation \cite{Chen2025}.

\subsection{Communication protocol}
FedKBP+ uses Google remote procedure call (gRPC) protocol \cite{Wang1993} as the communication stack \cite{Beutel2020}. gRPC is an open-source communication framework that operates over the Transmission Control Protocol (TCP) \cite{Kessler2004} using Hypertext Transfer Protocol Version 2 (HTTP/2) \cite{Belshe2015} as its transport layer. By exploiting key HTTP/2 features such as multiplexing, binary framing, and header compression, gRPC enhances communication efficiency and performance. TCP ensures a reliable connection, making gRPC well-suited for distributed systems. Its high throughput, low latency, and support for multiple programming languages have led to its widespread adoption across numerous FL frameworks, including OpenFL \cite{Foley2022}, NVFlare \cite{Roth2022}, and Flower \cite{Beutel2020}, etc.

\section{Experiments}
We comprehensively evaluated FedKBP+ across three predictive tasks in treatment planning (Figure \ref{fig_kbp+}): (1) OAR delineation, (2) tumor target segmentation, (3) 3D dose prediction. Each task was assessed using a dedicated public dataset to ensure reproducibility and relevance. Using data from the OpenKBP dataset \cite{Babier2021}, we compared the performance of centralized FedAvg \cite{McMahan2017} with both pooled and individual models under both IID and non-IID simulation settings. We further evaluated the performance of FedKBP+ against NVFlare on the task of brain tumor segmentation using 2021 BraTS challenge dataset \cite{Baid2021}. Finally, we evaluated the robustness of a decentralized FL strategy when a portion of participating sites randomly drop out at each round of federated training on the task of pancreas segmentation on MRI scans using PanSeg dataset  \cite{Zhang2024}. 
 
\subsection{Federated dose prediction}
In this study, we evaluate the effectiveness of the FedKBP+ framework for federated 3D dose prediction. Specifically, we compare the performance of the centralized FedAvg algorithm \cite{McMahan2017} against two baseline approaches: (1) Pooled training, where data from all sites is centrally aggregated; and (2) Individual training, where each site trains a local model independently on its local data. These comparisons are conducted under both independent and identically distributed (IID) and non-IID data distributions across participating sites to assess of impact of data heterogeneity on FL.

\subsubsection{Data}
This study utilized the publicly available dataset from the Open-Access Knowledge-Based Planning (OpenKBP) Grand Challenge \cite{Babier2021}. The OpenKBP dataset comprises 340 radiotherapy treatment plans for head and neck cancer, including 200 training cases, 40 validation cases, and 100 testing cases. The prescribed dose levels for these plans were 70, 63, and 56 Gy, each delivered in 35 fractions. The annotated OARs include the brainstem, right parotid, left parotid, spinal cord, larynx, mandible, and esophagus. To simulate a federated learning environment, the 200 training cases and 40 validation cases were randomly partitioned into 8 groups, each representing a site.

\begin{figure}[ht]
   \begin{center}
   \includegraphics[width=8cm]{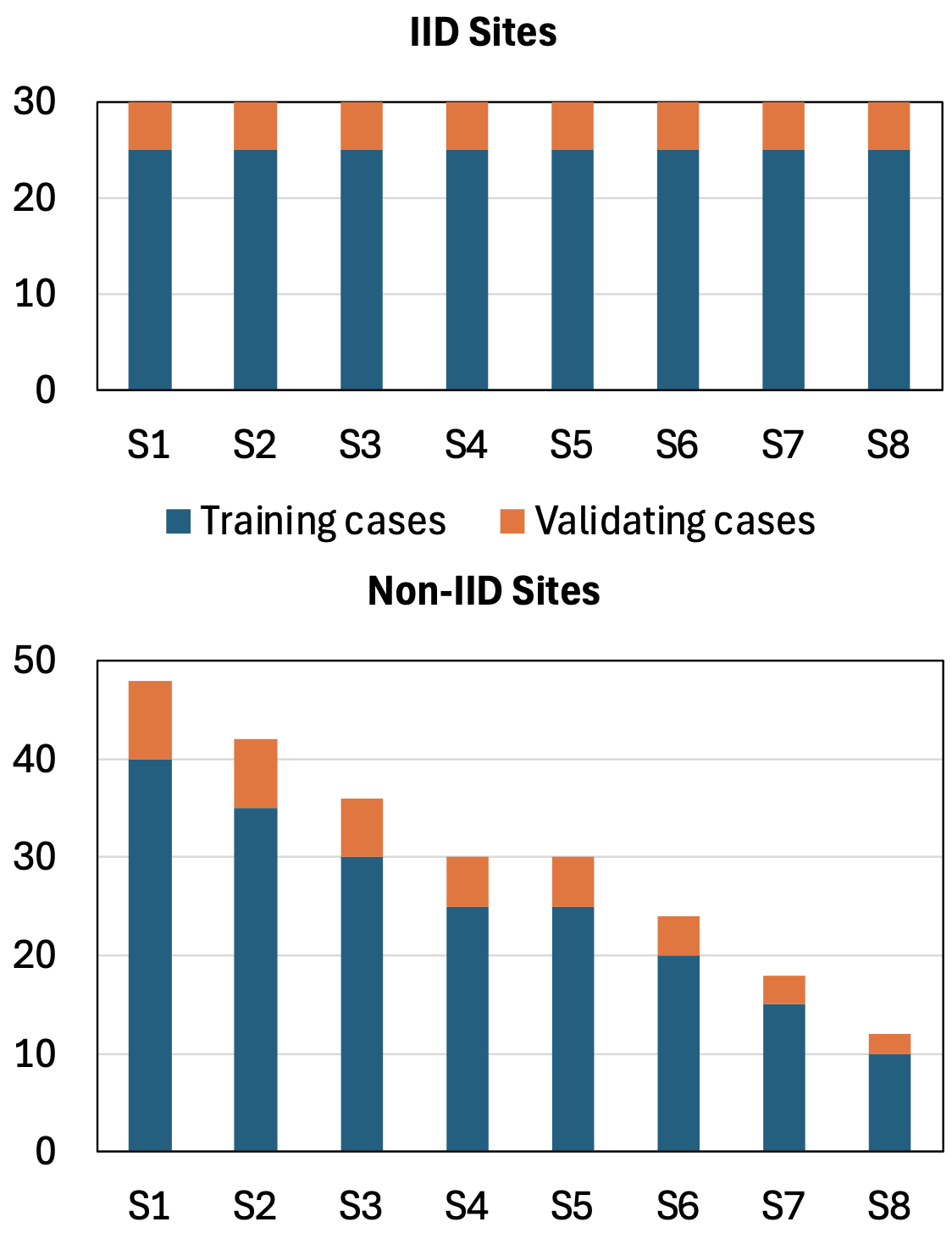}
   \captionv{16}{}
   {Case numbers for different sites of OpenKBP data. All sites share a common out-of-sample testing set of 100 cases (not shown in this figure). 
   \label{fig_split_openkbp} 
    }
    \end{center}
\end{figure}

To assess the impact of inter-site data heterogeneity on model performance, we considered two data distribution scenarios: (1) IID, where training and validation cases were evenly divided across all 8 sites; and (2) non-IID, where the number of cases varied across sites, introducing imbalance. The numbers of training and validating cases for different sites are visualized in Figure \ref{fig_split_openkbp}. The 100 testing cases were used as a common out-of-sample test set for all sites.

\subsubsection{FL settings}
In this case, we evaluated three distinct training approaches: (1) Pooled training, (2) Individual training, and (3) FedAvg \cite{Babier2021}, which constructs a global model at each round by computing a weighted average of the local models from participating sites. Given local training loss function $F_i$ and local model weights $w_i^t$ at round $t$, the global model weights at round $t+1$ can be computed with:

\begin{equation}
\label{eq_fedavg}
\begin{split}
\forall i, w_i^{t+1}=w_i^t-\eta_i\nabla F_i(w_i^t)\\
w^{t+1}=\sum_{i=1}^{N}\frac{m_i}{m}w_i^{t+1}
\end{split}
\end{equation}

For both FedAvg and Individual training, we further evaluated performance under two data distribution settings: IID and non-IID. Model performance was assessed using the dose and DVH scores, as defined in the OpenKBP Challenge \cite{Babier2021}. In both metrics, lower scores indicate higher prediction accuracy and overall better model performance.

\subsubsection{Implementations}
Within the FedKBP+ framework, each site is uniquely identified by a combination of its IP address and port number, enabling flexible deployment. Sites can be hosted on the same physical machine—using a shared IP with distinct ports—or distributed across multiple workstations with separate IP and port configurations. In this use case, an 8-site FedAvg setup was executed across eight NVIDIA GPUs housed in two workstations, with each site allocated a dedicated GPU to ensure parallel and isolated execution.

We utilized SA-Net as the backbone model, employing the voxel-wise Mean Absolute Error (MAE) between predicted and ground truth dose distributions \cite{Yuan2017,Adabi2022} as the loss function. All training methods were run for 100 epochs to ensure consistency in comparison.

\subsubsection{Results}
\begin{figure}[ht]
   \begin{center}
   \includegraphics[width=7cm]{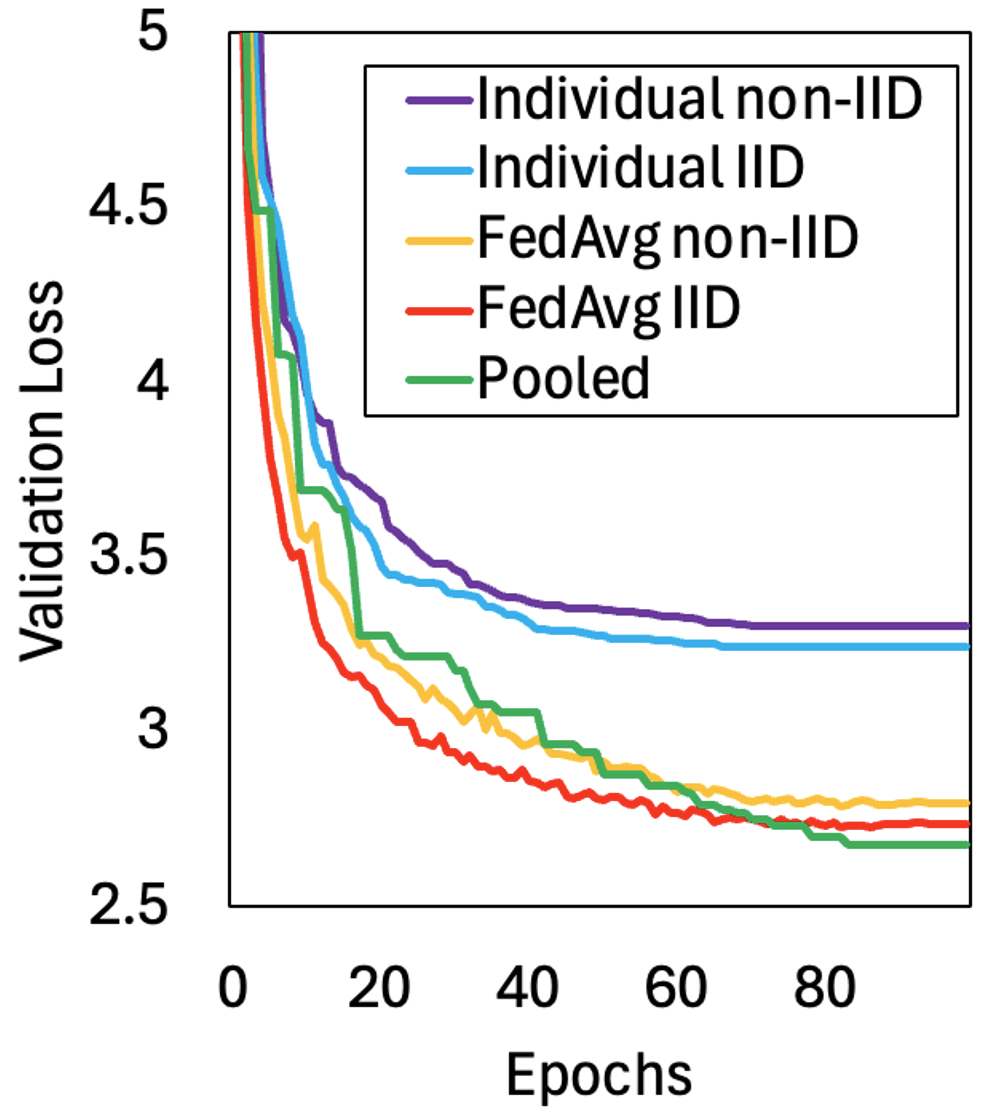}
   \captionv{16}{}
   {Validation loss over epochs for different training methods. For Individual training, the validation losses are averaged over 8 sites.
   \label{fig_valid_openkbp} 
    }
    \end{center}
\end{figure}

Figure \ref{fig_valid_openkbp} illustrates the validation loss over training epochs for the different training methods. For Individual training, the plotted curve represents the average validation loss across the 8 participating sites. As anticipated, Pooled training achieved the lowest final validation loss, benefiting from centralized data from different sites. The FedAvg approach, which facilitates model exchange at each epoch, achieved faster convergence and lower final validation losses compared to Individual training under both IID and non-IID settings. These findings underscore the effectiveness of FL in enhancing model optimization across distributed sites. Furthermore, in both FedAvg and Individual training, models trained on non-IID data distributions consistently lagged behind their IID counterparts, highlighting the challenges posed by data heterogeneity, which can impede both optimization efficiency and generalization in FL environments.

\begin{figure}[ht]
   \begin{center}
   \includegraphics[width=6cm]{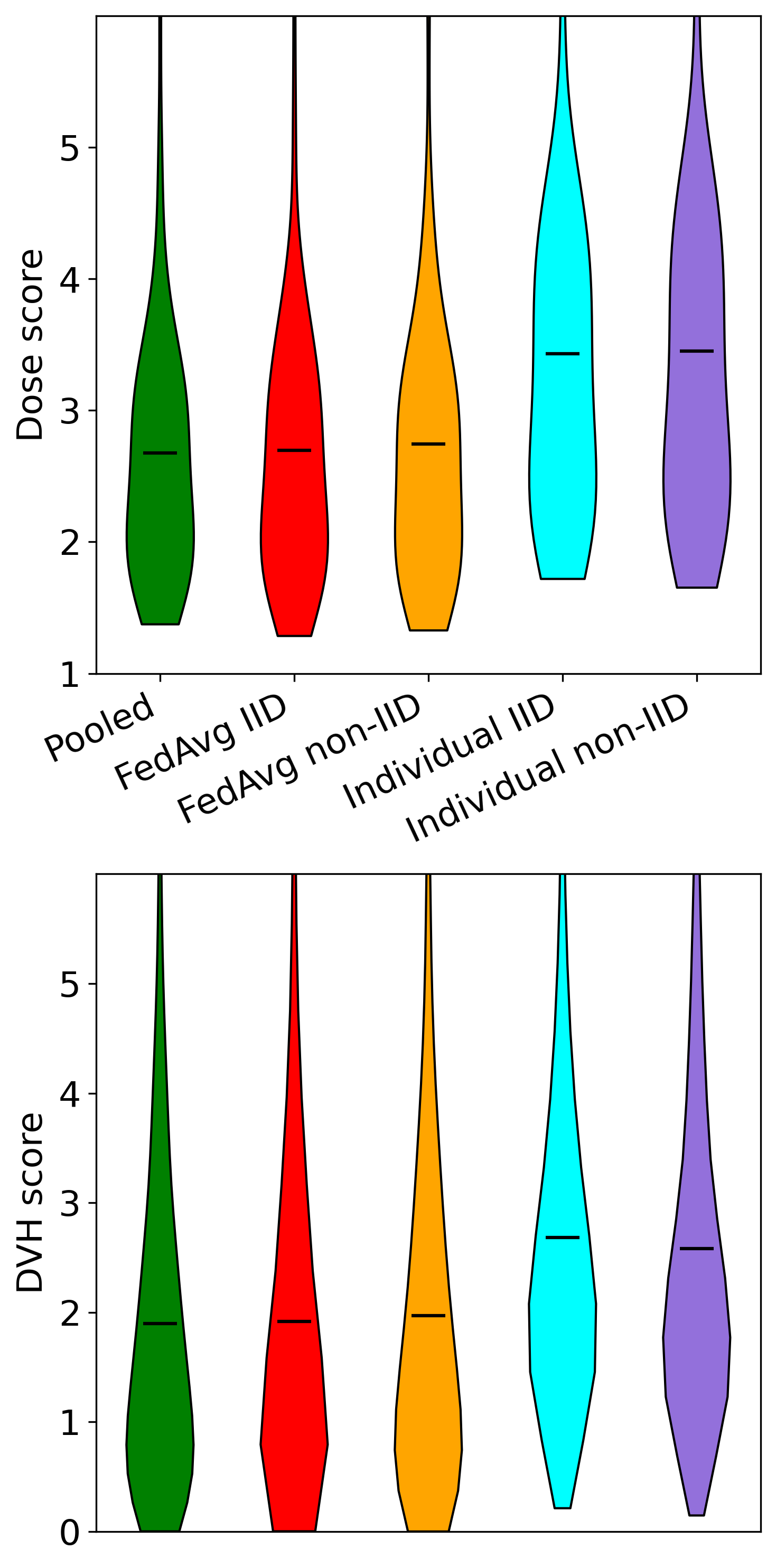}
   \captionv{16}{}
   {Violin plots of dose and DVH score distributions for different methods. The horizontal bars represent the mean values. For Individual training, the scores are averaged over 8 sites.
   \label{fig_test_global_openkbp} 
    }
    \end{center}
\end{figure}

Figure \ref{fig_test_global_openkbp} compares model performance of the different training methods on the testing dataset. For Individual training, performance metrics were averaged across all 8 sites. As expected, Pooled training achieved the best performance on both the dose score and DVH score, owing to its access to the fully aggregated dataset. FedAvg consistently outperformed Individual training across both evaluation metrics and under both IID (dose score: 2.6953 vs 3.4288; DVH score: 1.9196 vs 2.6816) and non-IID (dose score: 2.7452 vs 3.4528; DVH score: 1.9706 vs 2.5843) settings. Notably, IID FedAvg achieved performance levels comparable to Pooled training (dose score: 2.6953 vs 2.6758; DVH score: 1.9196 vs 1.8990), suggesting that FL can approach the performance of pooled-data training despite operating on distributed data. However, performance degradation was observed under non-IID FedAvg, which produced higher testing scores in both metrics compared to its IID counterpart (dose score: 2.7452 vs 2.6953; DVH score: 1.9706 vs 1.9196). These results are consistent with the validation loss trends observed in Figure \ref{fig_valid_openkbp}, underscoring the impact of data heterogeneity on model performance in federated settings.

\begin{figure}[ht]
   \begin{center}
   \includegraphics[width=15cm]{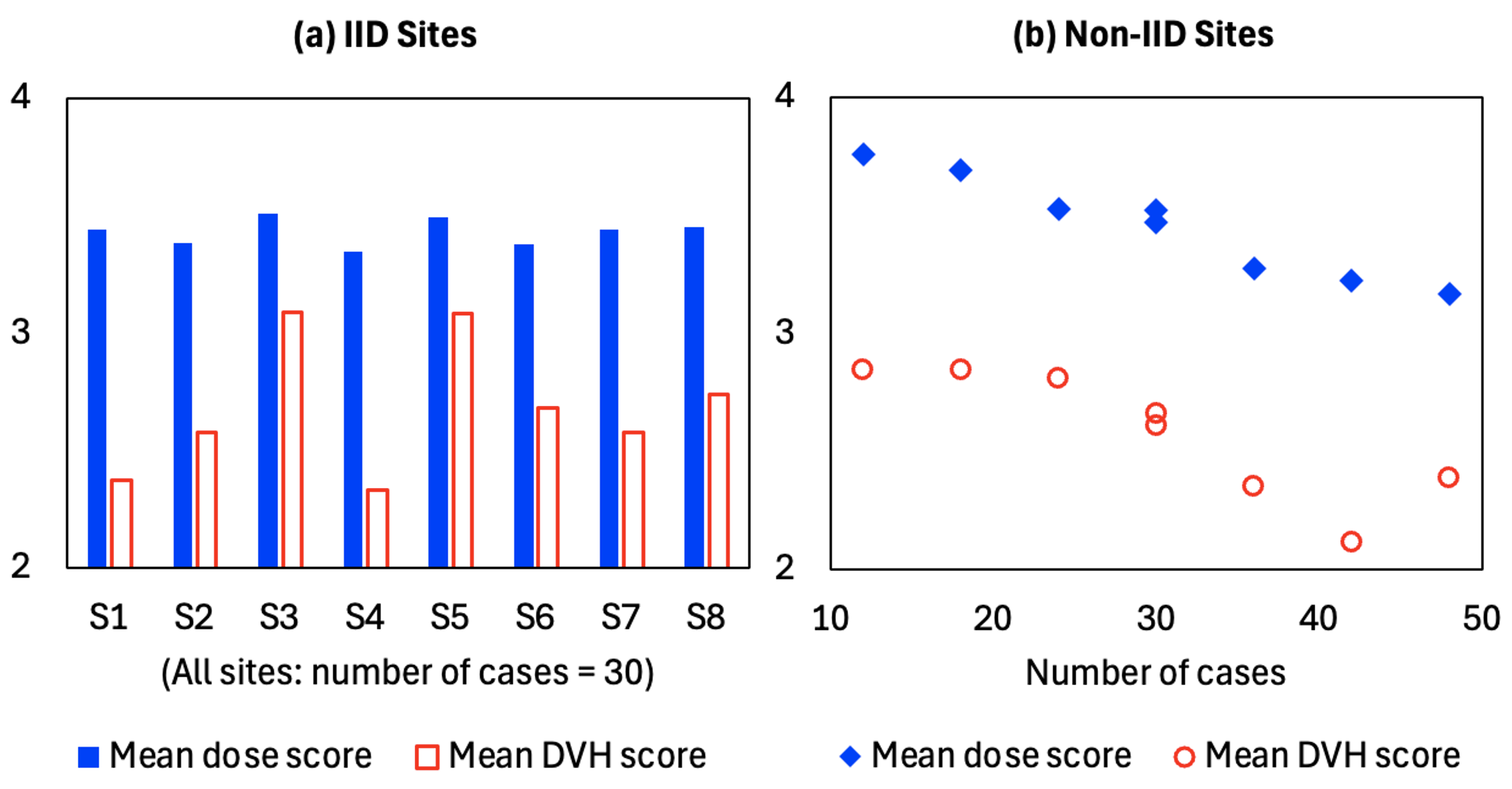}
   \captionv{16}{}
   {Mean dose and DVH scores of different sites under Individual training. Scores are showed as bars for IID sites(a) and as scatter points against case numbers for non-IID sites(b).
   \label{fig_test_local_openkbp} 
    }
    \end{center}
\end{figure}

To further investigate the impact inter-site data heterogeneity to FL, we compared the mean dose scores and DVH scores of different sites under Individual training. For the IID setting, site-level scores are visualized as bar plots (Figure \ref{fig_test_local_openkbp}(a)), while for the non-IID setting, they are displayed as scatter plots against the number of cases per site (Figure \ref{fig_test_local_openkbp}(b)). Under the IID distribution, site performances were generally consistent, with some degree of individual variability. In contrast, the non-IID setting revealed a clear relationship between dataset size and performance—larger sites consistently outperformed smaller ones. For example, Site 0, the largest non-IID site with 48 cases, outperformed Site 7, the smallest site with 12 cases, by approximately 18\% in dose score (3.1687 vs. 3.7557) and 19\% in DVH score (2.3872 vs. 2.8450). These findings highlight the advantage of larger local datasets and reinforce the importance of collaboration among data contributors in federated settings to achieve improved model performance across sites.

A key advantage of the FedKBP+ framework is its ability to leverage parallel processing, which substantially reduces overall training time. For instance, FedAvg training under IID conditions was completed in just 13.37 hours using FedKBP+, compared to 86.21 hours required for sequential site-by-site training—demonstrating a significant gain in efficiency.

Due to the absence of site information in the OpenKBP dataset, we simulated a non-IID data distribution by unevenly distributing cases across sites. In contrast, the following two use cases employed datasets containing explicit site identifiers, enabling us to partition the data based on actual institutional sources. This allowed for the creation of a more realistic non-IID environment, providing a stronger foundation for evaluating FL performance.

\subsection{Federated tumor target segmentation}
In this study, we applied FedKBP+ for tumor target segmentation using the publicly available training data from the Brain Tumor Segmentation (BraTS) Challenge \cite{Baid2021}. The BraTS Challenge offers a standardized platform for developing and evaluating automated segmentation algorithms for brain tumors using multi-parametric MRI scans, facilitating consistent and fair comparisons across methods in segmenting key tumor sub-regions. To further validate the performance of FedKBP+, we conducted a comparative evaluation against the open-source NVFlare framework \cite{Roth2022}.

\subsubsection{Data}
\begin{figure}[ht]
   \begin{center}
   \includegraphics[width=9cm]{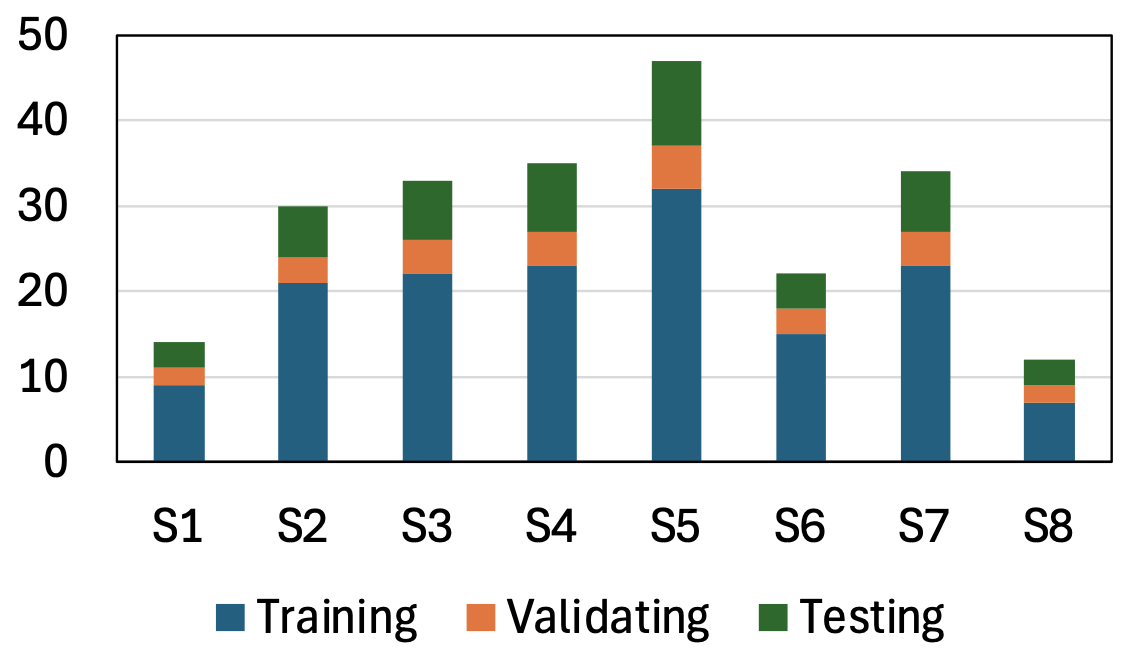}
   \captionv{16}{}
   {Case numbers for different sites of BraTS data. 
   \label{fig_split_brats} 
    }
    \end{center}
\end{figure}

We used 227 multi-parametric MRI cases from 8 sites in the BraTS 2021 Challenge dataset. Each case includes four MRI modalities: native T1-weighted, post-contrast T1-weighted, T2-weighted, and T2 Fluid-Attenuated Inversion Recovery (FLAIR) images. The dataset provides detailed annotations for three tumor sub-regions: enhancing tumor, necrotic and non-enhancing tumor core, and peritumoral edema. Within each site, the data was split into approximately 70\% for training, 10\% for validation, and 20\% for testing, as illustrated in Figure \ref{fig_split_brats}.

\subsubsection{FL settings}
In additional to the baseline approaches of Pooled and Individual training, we implemented two centralized FL algorithms in both FedKBP+ and NVFlare frameworks: (1) FedAvg \cite{McMahan2017} and (2) FedProx. FedProx is a variant of FedAvg that incorporates a proximity term into the loss function to better handle data heterogeneity across sites \cite{Li2018}. Given local training loss function $F_i$, local model weights $w_i^t$, and global model weights $w^t$ at round $t$, the global model in FedProx can be updated with:

\begin{equation}
\label{eq_fedprox}
\begin{split}
\forall i, \widetilde{F}(w_i^t)=F_i(w_i^t)+\frac{\mu}{2}\left\| w_i^t-w^t \right\|^2\\
\forall i, w_i^{t+1}=w_i^t-\eta_i\nabla \widetilde{F}_i(w_i^t)\\
w^{t+1}=\sum_{i=1}^{N}\frac{m_i}{m}w_i^{t+1}
\end{split}
\end{equation}

where the proximity term $\left\| w_i^t-w^t \right\|^2$ regulates the difference between local model and global model at each round, and $\mu$ controls the balance between loss function $F_i$ and the proximity term. Model performance was assessed using the mean Dice Similarity Coefficient (DSC) computed on the test data.

\subsubsection{Implementations}
Both  FedKBP+ and NVFlare  were deployed across 8 NVIDIA GPUs  over two workstations, with each federated site assigned a dedicated GPU to perform local training. All methods were trained for 400 epochs to ensure a consistent evaluation. NVFlare uses the winner  of the BraTS 2018 Challenge \cite{Myronenko2018} as the segmentation model, while FedKBP+ employed SA-Net that achieved the 2nd place in the BraTS 2021 challenge\cite{Yuan2022} (code available at: \url{https://github.com/CUMC-Yuan-Lab/GCML/tree/main/SANet/BraTS}). SA-Net was optimized using a hybrid loss function that combines Jaccard distance and voxel-wise focal loss to enhance segmentation accuracy and robustness.

\subsubsection{Results}
\begin{figure}[ht]
   \begin{center}
   \includegraphics[width=7cm]{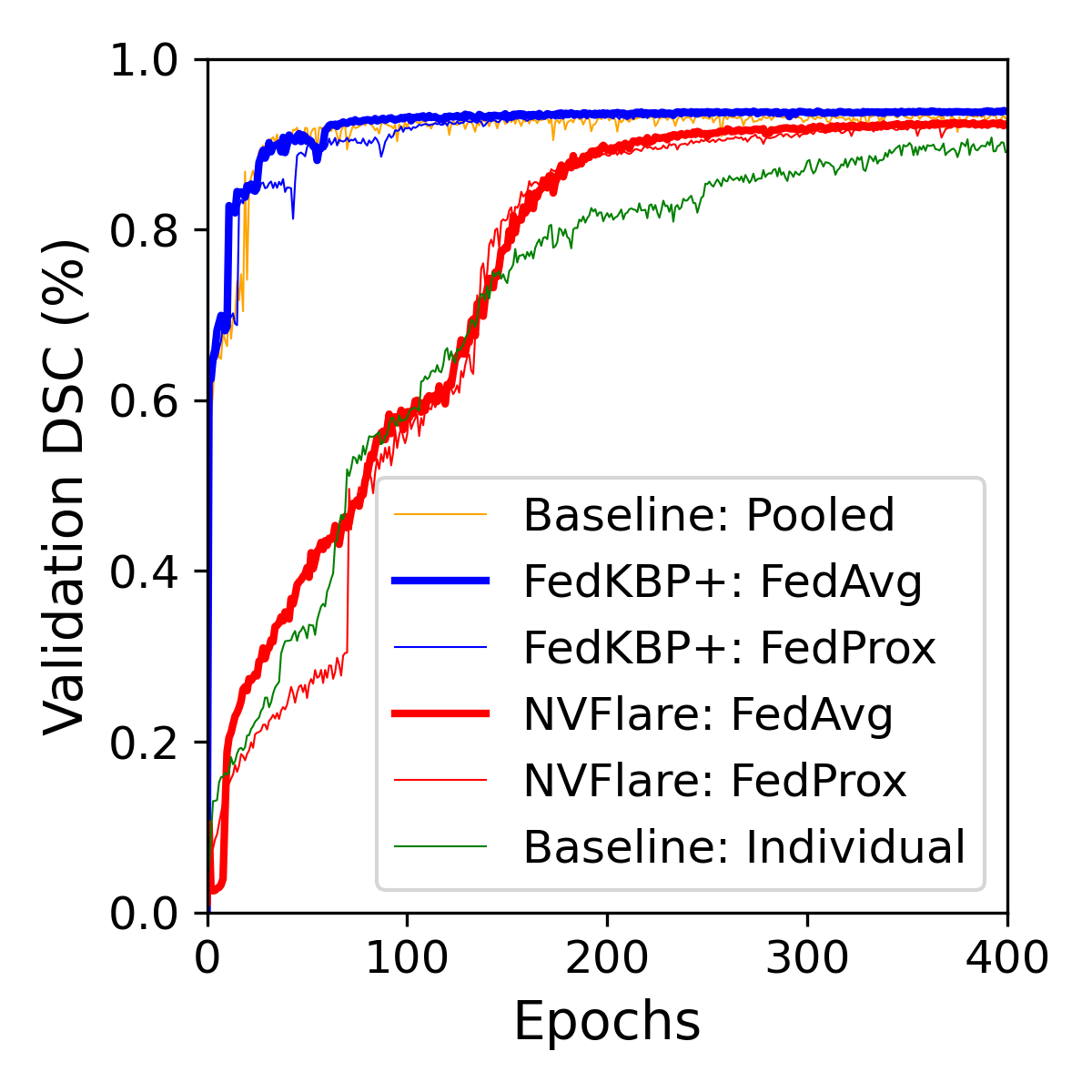}
   \captionv{16}{}
   {Change of validation DSC over epochs for different training methods. FedKBP+ implementations are marked in blue (thick line: FedAvg; thin line: FedProx). Likewise, NVFlare implementations are in red. For Individual training, DSC is averaged over 8 sites.
   \label{fig_valid_brats} 
    }
    \end{center}
\end{figure}

The progression of validation DSC over training epochs is shown in Figure \ref{fig_valid_brats}, reflecting the model optimization dynamics over time. Both FL methods (FedAvg and FedProx) outperformed Individual training in final DSC, regardless of framework. In both frameworks, FedProx exhibited slower convergence compared to FedAvg, which is expected due to the presence of a proximity term in its loss function that adds regularization to account for inter-site heterogeneity. Despite this, the final model performance of FedProx was comparable to FedAvg after convergence.

When comparing the two frameworks, FedKBP+ consistently demonstrated much faster convergence and achieved higher DSC values than the NVFlare baseline across both FL algorithms. This performance gain is primarily attributed to the advanced scale attention mechanism embedded in SA-Net \cite{Yuan2021}, which enhances feature representation across multiple scales. Notably, the FedKBP+ implementations of both FedAvg and FedProx achieved final DSC performance comparable to the Pooled training baseline, further demonstrating the potential of federated learning as a viable alternative to centralized model training on aggregated data.

\begin{figure}[ht]
   \begin{center}
   \includegraphics[width=12cm]{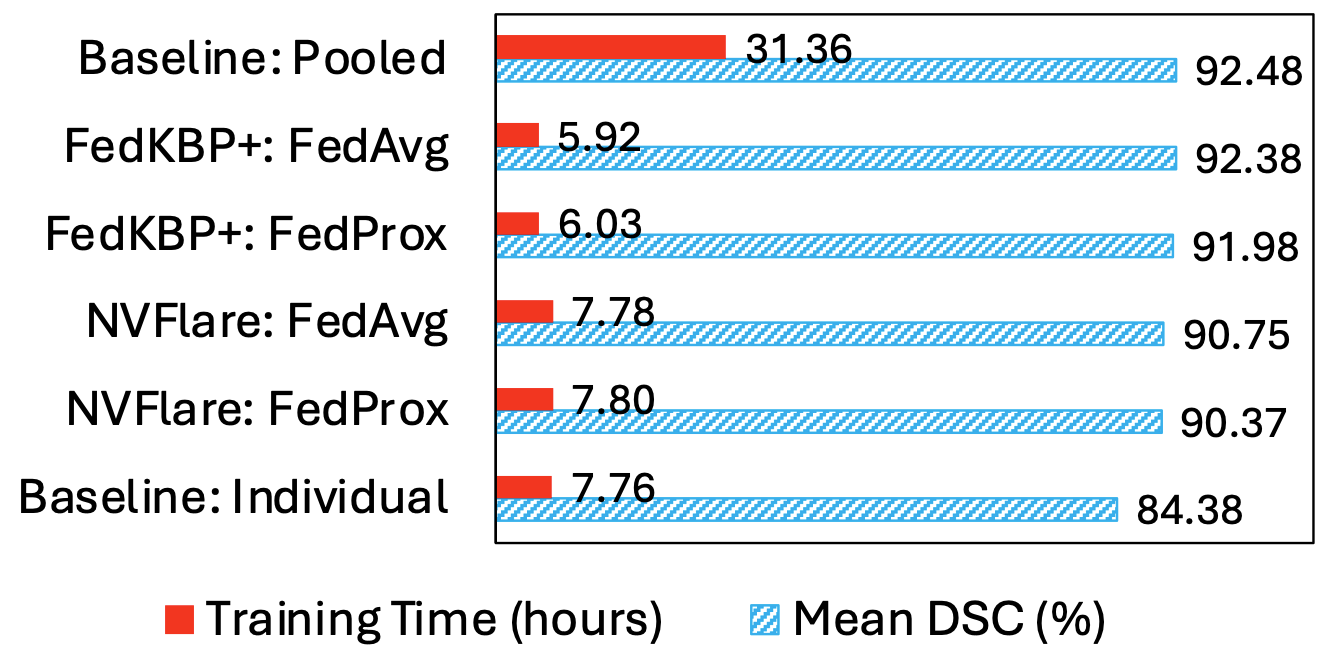}
   \captionv{16}{}
   {Compare of model training time and mean DSC for different methods. For the baseline of Individual, model performance and training time are averaged 8 sites.
   \label{fig_test_brats} 
    }
    \end{center}
\end{figure}

Figure \ref{fig_test_brats} presents the resulting DSC on testing data and the total training time for different methods. For the Individual training, both DSC and training time were averaged across the 8 participating sites. As expected, Pooled training achieved the highest DSC, but also incurred the longest training time due to the centralized processing of the entire dataset. Both FL methods (FedAvg and FedProx) showed improved model performance over the Individual training. Among them, FedAvg consistently outperformed FedProx in terms of both accuracy and efficiency, across both FL frameworks. Specifically, FedAvg achieved higher DSC  than FedProx (FedKBP+: 92.38\% vs 91.98\%; NVFlare: 90.75\% vs 90.37\%), and required less training time in hours (FedKBP+: 5.92 vs 6.03; NVFlare: 7.78 vs 7.80).

When comparing the two frameworks, FedKBP+ consistently outperformed NVFlare, achieving higher DSC scores (FedAvg: 92.38\% vs 90.75\%; FedProx: 91.98\% vs 90.37\%) and substantially shorter training time (FedAvg: 5.92 vs 7.78; FedProx: 6.03 vs 7.80). These findings underscore FedKBP+ as an efficient FL framework with high performance in predictive tasks, aligning with the validation trends observed in Figure \ref{fig_valid_brats}.

\subsection{Federated OAR segmentation}

In this study, we applied Gossip Contrast Mutual Learning (GCML), a decentralized FL strategy, for OAR segmentation using the recently released PanSeg dataset. As illustrated in Figure \ref{fig_decentralized_fl}, decentralized FL operates without reliance on a central server for model exchange. Instead, it functions in a fully distributed manner through direct P2P communication \cite{Rieke2020}, offering potential advantages in both model performance \cite{Chen2025, Li2022} and security \cite{He2022, Che2022}.  Our previous study has demonstrated that GCML achieved superior performance compared to several state-of-the-art centralized and decentralized FL approaches in various medical image segmentation tasks \cite{Chen2025}, therefore, this study will focus on the robustness of GCML within the FedKBP+ framework under dynamic training conditions involving random site drop-in and drop-out, which are representative of real-world operational environments.

\subsubsection{Data}

\begin{figure}[ht]
   \begin{center}
   \includegraphics[width=8cm]{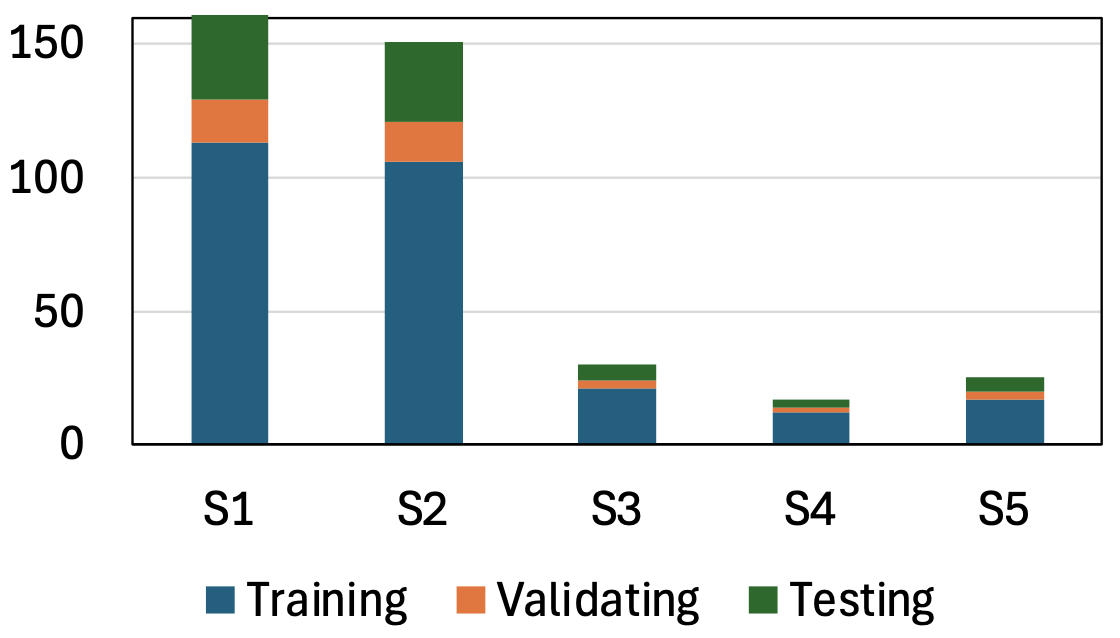}
   \captionv{16}{}
   {Case numbers for different sites of PanSeg data. 
   \label{fig_split_panseg} 
    }
    \end{center}
\end{figure}

While OAR delineation is typically performed on CT images, with recent advances in MRI-LINAC technology \cite{Alnaghy2020}, MRI-based OAR delineation has become increasingly viable and clinically relevant. However, automated OAR segmentation on MRI remains understudied. To explore this potential, we evaluated  FedKBP+ framework using the PanSeg dataset \cite{Zhang2024}, the largest publicly available MRI dataset dedicated to automated pancreas segmentation. This study utilized 384 T1-weighted scans collected from five contributing institutions. For each institution, the data was partitioned into approximately 70\% for training, 10\% for validation, and 20\% for testing, as illustrated in Figure \ref{fig_split_panseg}.

\subsubsection{FL settings}

GCML algorithm \cite{Chen2025} employs gossip protocol \cite{Baraglia2013} for P2P model exchange in a fully decentralized FL environment. To further improve the utility of peer models, GCML introduces Deep Contrastive Mutual Learning (DCML) that leverages a contrastive learning mechanism in which both the local and incoming models are guided to align their predictions at voxels where a reference model yields correct classifications, and to diverge from the reference model at voxels where it provides the wrong prediction. This strategy promotes effective mutual knowledge distillation while mitigating the spread of erroneous information. Our previous study  on different segmentation tasks using three publicly available datasets showed that GCML framework outperformed both centerlized and decentralzied FL methods with significantly reduced communication overhead \cite{Chen2025}. 

Within the FedKBP+ framework, GCML is executed via the procedure outlined in Algorithm \ref{alg_p2p}. The key step is the regional DCML, which facilitates knowledge transfer between local and received models within the region of interest on local data. With $P_r^t$ and $P_s^t$ denoting the predicted probability distribution of local model $w_r^t$ (on Receiver site) and incoming model $w_s^t$ (from Sender site) at round $t$, the local model at round $t+1$ can be updated with:

\begin{equation}
\label{eq_dcml}
\begin{split}
\widehat{F}_r(w_r^t) = (1 - \lambda) F_r(w_r^t) + \lambda D_{CKL}(P_r^t \parallel P_s^t)\\
\widehat{F}_s(w_s^t) = (1 - \lambda) F_r(w_s^t) + \lambda D_{CKL}(P_s^t \parallel P_r^t)\\
 w_r=w_r^t-\eta_r\nabla \widehat{F}_r(w_r^t)\\
 w_s=w_s^t-\eta_r\nabla \widehat{F}_s(w_s^t)\\
 w_r^{t+1}=\frac{v_rw_r+v_sw_s}{v_r+v_s}
\end{split}
\end{equation} 

where $F_r$ is the local training loss, $D_{CKL}$ is the regional contrastive K-L Divergence \cite{Chen2025}, $\lambda$ is an parameter to balance the two loss terms, $v_r$ and $v_s$ are the validation loss of model $w_r$ and $w_s$ respectively. Note $F_r$ is used for both $\widehat{F}_r$ and $\widehat{F}_s$ as both models are trained on site $r$. For more details, please refer to our previous work \cite{Chen2025}. 

\begin{algorithm}
\caption{GCML in FedKBP+}\label{alg_p2p}
\begin{algorithmic}
\State \textbf{Inputs:} 
\\\hspace*{4mm} $R_{total}$: total number of FL rounds
\\\hspace*{5mm}$R_{current}$: current number of FL rounds
\While{$R_{current}<=R_{total}$}
\State \textbf{Coordinator side:}
\\\hspace*{12mm} Receive status update from sites and form a list of active sites
\\\hspace*{12mm} Select pairs of Sender/Receiver sites from the list of active sites
\\\hspace*{12mm} Broadcast the addresses (IP:Port) and IDs of Sender and Receiver sites
\State \textbf{Site side:}
\\\hspace*{12mm} Receive broadcast from coordinator
\\\hspace*{12mm} Check membership to the Sender and Receiver lists
\\\hspace*{12mm} \textbf{if} site is a Sender \textbf{then}
\\\hspace*{20mm} Look up the address (IP:Port) of corresponding Receiver site 
\\\hspace*{20mm} Send local model to Receiver site
\\\hspace*{12mm} \textbf{else if} site is a Receiver \textbf{then}
\\\hspace*{20mm} Receive incoming model from Sender site 
\\\hspace*{20mm} Conduct regional DCML and update local model
\\\hspace*{12mm} Conduct local training
\\\hspace*{12mm} Send status update to coordinator
\EndWhile
\end{algorithmic}
\end{algorithm}

Algorithm \ref{alg_drop_in_out}  list the simulation procedure for the unavailability of a predefined portion of participating sites during FL in order to systematically evaluate the impact of random site drop-in and drop-out under controlled conditions. We evaluated the model performance and total training time when  $N_{max}=0$, 1 and 2 respectively, representing up to 0\%, 20\% and 40\% of drop-out sites. For $N_{max}>0$ , we simulated two distinct scenarios: (1) Local training on drop-out sites – simulating temporary internet disconnection, where local training continues but model updates are not communicated; (2) No local training on dropped-out sites – simulating workstation shutdown, where both local training and communication are suspended.

\begin{algorithm}
\caption{Simulation of random site drop in and out during FL}\label{alg_drop_in_out}
\begin{algorithmic}
\State \textbf{Inputs:} 
\\\hspace*{4mm} $R_{total}$: total number of FL rounds
\\\hspace*{5mm}$R_{current}$: current number of FL rounds
\\\hspace*{5mm}$N_{total}$: total number of sites
\\\hspace*{5mm}$N_{current}$: current number of sites
\\\hspace*{5mm}$N_{max}$: max number of drop-out sites
\State \textbf{Initialize:} $N_{current} \gets N_{total}$
\While{$R_{current}<=R_{max}$}
\If{$N_{current}=N_{total}$}
    \State $1/2$ chance of one site drop out: $N_{current} \gets N_{current}-1$
    \State $1/2$ chance of no site drop in or out
\ElsIf{$N_{current}=N_{total}-N_{max}$}
    \State $1/2$ chance of one site dropping in: $N_{current} \gets N_{current}+1$
    \State $1/2$ chance of no site dropping in or out
\Else
    \State $1/3$ chance of one site drop out: $N_{current} \gets N_{current}-1$
    \State $1/3$ chance of one site drop in: $N_{current} \gets N_{current}+1$
    \State $1/3$ chance of no site drop in or out
\EndIf
\EndWhile
\end{algorithmic}
\end{algorithm}

\subsubsection{Implementations}
The 5-site GCML training was conducted using 5 NVIDIA GPUs distributed over two workstations, with each federated site assigned a dedicated GPU to perform local training. Each drop-in/out scenario was trained for 500 epochs to ensure consistency across evaluations. SA-Net was used as the backbone model for pancreas segmentation, employing a composite loss function that combines cross-entropy loss \cite{Krause2023} with Jaccard distance \cite{Yuan2017b} to balance pixel-wise classification and overlap-based accuracy.

\subsubsection{Results}
\begin{figure}[ht]
   \begin{center}
   \includegraphics[width=16cm]{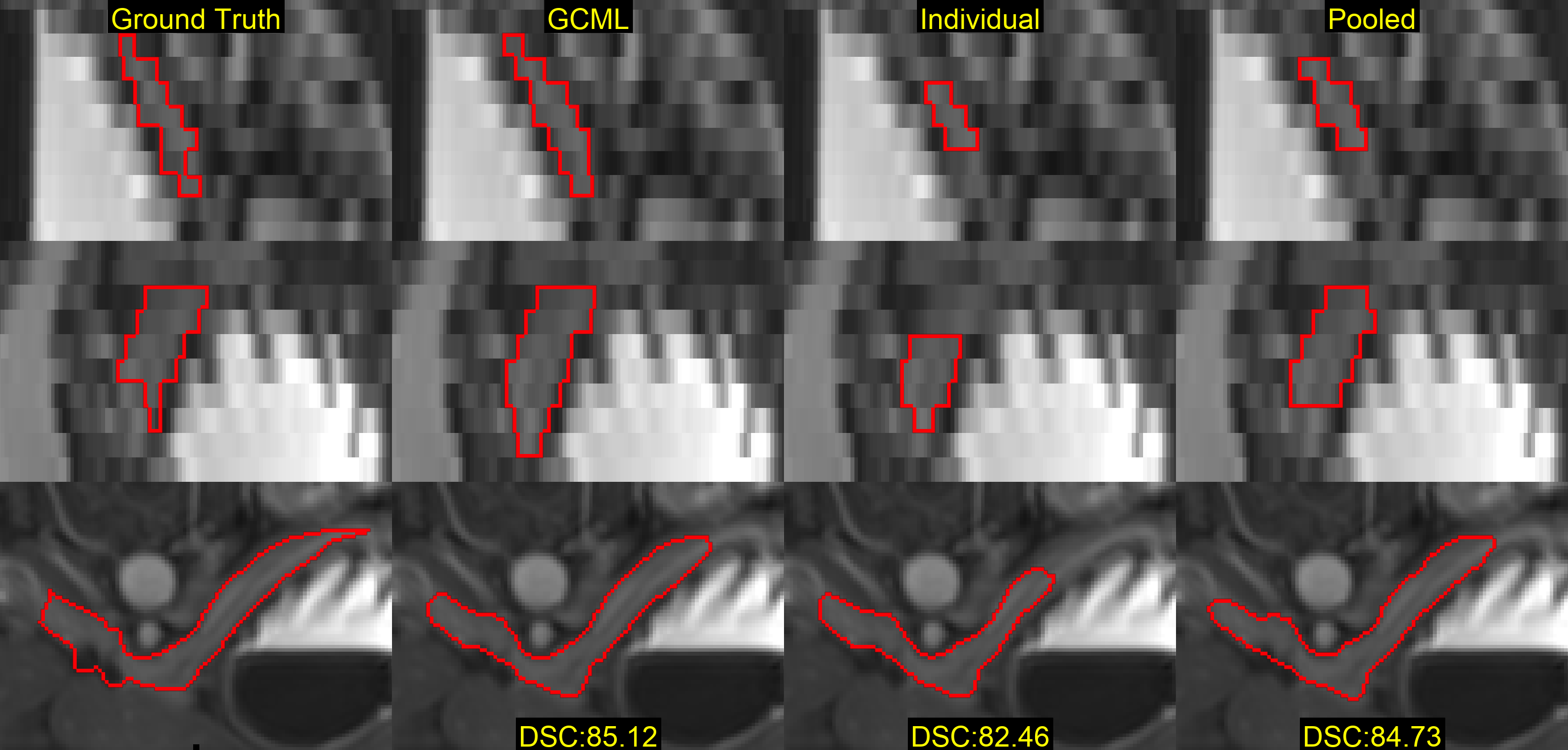}
   \captionv{16}{}
   {Compare of segmentation masks and DSCs of different training methods on a typical case. Top row: sagittal view; middle: coronal; bottom: axial.    
   \label{fig_mask_panseg} 
    }
    \end{center}
\end{figure}

Our previous study has shown that GCML delivers superior model performance compared to traditional centralized FL methods such as FedAvg and FedProx, and achieves accuracy that is comparable to or better than the Pooled training \cite{Chen2025}. These performance advantages are consistently reproduced within the FedKBP+ framework. As illustrated in Figure \ref{fig_mask_panseg}, GCML yielded better segmentation results as compared to the individual model, demonstrating better agreement with the ground truth and achieving quality comparable to the pooled model. These findings underscore GCML’s effectiveness in delivering high-quality segmentation model within a fully decentralized federated learning framework.

Figure \ref{fig_drop_panseg} lists model performance under different drop-in/out settings. Model performance was only marginally impacted by the reduced participating sites during training. An ANOVA test on the DSCs of individual cases revealed no statistically significant differences among the five scenarios ($p=0.9097$). These results underscore the robustness of the GCML algorithm and highlight the viability of decentralized federated learning as a resilient and practical approach for real-world clinical applications.

\section{Discussion}
Predictive models are playing increasingly important roles in treatment planning practice. In this study, we introduced  KBP+ to integrate the prediction tasks of OAR delineation, tumor segmentation, and 3D dose distribution into a unified machine learning pipeline. This unified approach enables a streamlined workflow, promoting more efficient and cohesive processing across the radiotherapy planning pipeline.

\begin{figure}[ht]
   \begin{center}
   \includegraphics[width=7cm]{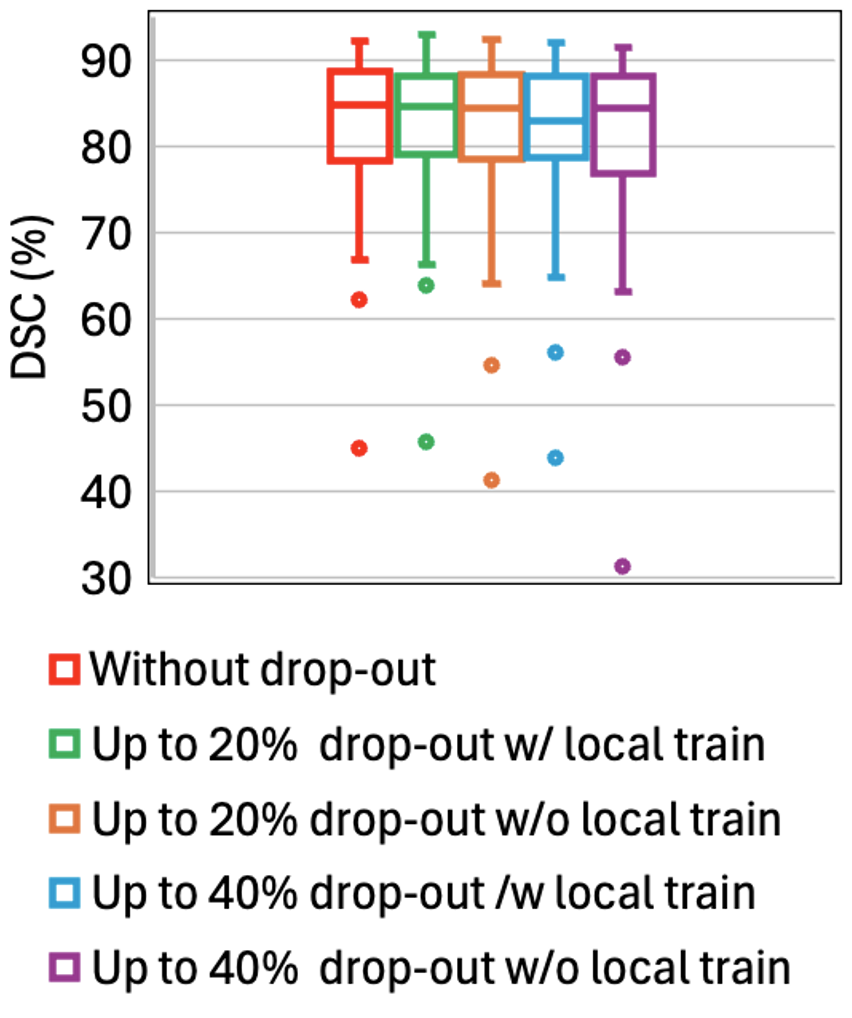}
   \captionv{16}{}
   {Box plot of DSCs with different percentages of site drop-out 
   \label{fig_drop_panseg} 
    }
    \end{center}
\end{figure}

All three predictive tasks encompassed within the KBP+ framework stand to benefit from FL, which facilitates collaborative model training across multiple institutions while safeguarding the privacy of sensitive patient information. Our results demonstrate that FL outperforms isolated local training and achieves performance on par with pooled-data training across all predictive tasks, as evidenced by the results presented in Figures \ref{fig_test_global_openkbp}, \ref{fig_valid_brats}, \ref{fig_test_brats}, and \ref{fig_mask_panseg}.

Despite the growing promise of FL in healthcare, real-world medical FL deployments face substantial practical challenges, including inter-site data heterogeneity and the unpredictability of site participation due to random drop-in and drop-out events \cite{Rieke2020, Li2018}. FedKBP+ is specifically designed to address these challenges by providing a robust, secure, and flexible FL framework tailored to the tasks in radiation therapy. To the best of our knowledge, FedKBP+ represents the first deployment-ready FL platform designed for predictive tasks in RT. Compared to existing open-source FL frameworks, FedKBP+ offers the following key advantages:

\textbf{Efficiency and Effectiveness:} FedKBP+ demonstrates superior training efficiency and model performance compared to existing open-source FL frameworks. In the brain tumor segmentation use case, FedKBP+ (with SA-Net backbone) outperformed NVFlare in both model accuracy and total training time (FedAvg: 92.38\% in 5.92 hours vs 90.75\% in 7.78 hours; FedProx: 91.98\% in 6.03 hours vs 90.37\% in 7.80 hours), evident in Figure \ref{fig_test_brats}. This superiority of GCML is further underscored by the validation DSC trajectories (Figure \ref{fig_valid_brats}), which highlight the efficacy of SA-Net’s embedded scale attention mechanism (Figure \ref{fig_sanet}).

\textbf{Flexible Configuration:} FedKBP+ supports both centralized (Figure \ref{fig_centralized_fl}) and decentralized (Figure \ref{fig_decentralized_fl}) FL paradigms, offering exceptional architectural flexibility. Traditional federated learning algorithms, such as FedAvg \cite{McMahan2017} and FedProx \cite{Li2018}, are based on a centralized client–server architecture. While this configuration is effective in many scenarios, it inherently introduces a single point of failure, as the system's operation is critically dependent on the availability and reliability of the central server. By contrast, decentralized FL eliminates this reliance, enhancing robustness and adaptability. However, decentralized approaches typically employ P2P communication, which constrains information exchange to local subsets of participants, potentially limiting convergence speed and model generalization. To address this, FedKBP+ integrates GCML \cite{Chen2025}, an advanced algorithm that enhances knowledge propagation in decentralized settings. GCML combines contrastive mutual learning with gossip protocol to facilitate effective model updates. Prior studies have shown that decentralized FL, when properly optimized, can surpass centralized approaches in heterogeneous environments \cite{Chen2025, Li2022}.

\textbf{Robustness:} In real-world FL, site availability is inherently volatile due to unstable network connections or system interruptions. FedKBP+ mitigates this issue through the integration of GCML, which dynamically pairs participants in each communication round. This pairing strategy ensures continual learning progression, even with inconsistent site participation. Moreover, GCML’s contrastive learning mechanism improves model robustness by aligning predictions where peer models agree and encouraging divergence where errors are identified. As shown in Figure \ref{fig_drop_panseg}, FedKBP+ maintains high model performance despite up to 40\% site drop-out, demonstrating its resilience in dynamic operational settings.

\textbf{Portability and Scalability:} FedKBP+ features a modular and extensible architecture (Figures \ref{fig_centralized_fl} and \ref{fig_decentralized_fl}) that promotes ease of customization across a range of clinical applications. Core components—including model architectures, loss functions, training routines, and communication protocols—are independently configurable, allowing researchers to tailor the system to specific predictive tasks or clinical workflows. In addition, FedKBP+ incorporates task-agnostic scripting, which decouples FL logic from domain-specific model development. Users can readily extend the framework to new datasets or predictive objectives by substituting in compatible models and data loaders, without modifying the core infrastructure.

Together, these features establish FedKBP+ as a robust and adaptable solution for real-world FL deployments in RT planning. Its scalability enables efficient management of growing data volumes and expanding institutional participation, while its modular and portable architecture supports seamless adaptation to diverse anatomical targets, imaging modalities, and clinical objectives. By addressing core challenges such as data heterogeneity, system resilience, and communication flexibility, FedKBP+ provides a comprehensive foundation for advancing privacy-preserving, collaborative AI development in \textbf{RT-specific} predictive modeling.

One limitation of this study lies in the use of the OpenKBP dataset for dose prediction. While suitable for proof-of-concept purposes, the dataset is limited to a relatively small sample size and lacks metadata specifying the originating institutions or centers. As a result, the FL experiments based on this dataset may not fully capture the complexities and heterogeneity typical of real-world FL scenarios. To enable more realistic and generalizable investigations into federated dose prediction, future studies will benefit from access to larger, more diverse datasets that include detailed source-site annotations.

\section{Conclusions}
In this paper, we present FedKBP+, a comprehensive federated learning framework that integrates several predictive tasks in treatment planning into a unified collaborative learning pipeline. Our results clearly demonstrate that FedKBP+ is highly effective, efficient and robust, showing great potential as a federated learning platform for radiation therapy. 

\section*{Acknowledgments}
This work is supported by a research grant from Varian Medical Systems (Palo Alto, CA, USA), UL1TR001433 from the National Center for Advancing Translational Sciences, and R21EB030209 from the National institutes of Biomedical Imaging and Bioengineering of the National Institutes of Health, USA. The content is solely the responsibility of the authors and does not necessarily represent the official views of the National Institutes of Health. This research has been partially funded through the generous support of Herbert and Florence Irving/the Irving Trust.

\section*{Conflict of Interest Statement}
The authors have no relevant conflicts of interest to disclose.

\section*{References}
\addcontentsline{toc}{section}{\numberline{}References}
\vspace*{-20mm}
\bibliography{mendeley}      
\bibliographystyle{medphy.bst}    

\end{document}